\documentstyle[12pt]{article}
\begin {document}

\begin{flushright}
UT-598
\\
March  1992
\end{flushright}
\begin{center}
{\large\bf Topological Gauged WZW Models   \\
      and 2D Gravity } \vspace{.7 in}\\
{\bf Toshio Nakatsu and
        Yuji Sugawara \vspace{.5 in}\\
        {\it Department of Physics, University of Tokyo}\\
        {\it Bunkyo-ku,Tokyo 113,Japan \vspace{.8 in}}}
\end{center}

\newcommand{\br}{{\bf R}}
\newcommand{\bc}{{\bf C}}
\newcommand{\bz}{{\bf Z}}

\newcommand{\Om}{\Omega}

\newcommand{\gerg}{\mbox{g}}

\newcommand{\germ}{\mbox{m}}
\newcommand{\om}{\omega}
\newcommand{\al}{\alpha}
\newcommand{\la}{\lambda}

\newcommand{\De}{\Delta}
\newcommand{\vphi}{\varphi}

\newcommand{\gerh}{\mbox{h}}

\newcommand{\gerA}{{\cal A}}

\newcommand{\bm}[1]{\mbox{\boldmath ${#1}$}}
\newcommand{\ca}[1]{{\cal #1}}
\newcommand{\ger}[1]{\mbox{#1}}
\newcommand{\lb}{\lbrack}
\newcommand{\rb}{\rbrack}

\newcommand{\msc}[1]{\mbox{\scriptsize #1}}
\newcommand{\dsp}{\displaystyle}
\newcommand{\scs}[1]{{\scriptstyle #1}}
\newcommand{\deebar}{\bar{\partial}}

\newcommand{\llerarr}{\longleftrightarrow}
\newcommand{\Llerarr}{\Longleftrightarrow}

\newcommand{\be}{\begin{equation}}\newcommand{\ee}{\end{equation}}
\newcommand{\bea}{\begin{eqnarray}} \newcommand{\eea}{\end{eqnarray}}
\newcommand{\ba}[1]{\begin{array}{#1}} \newcommand{\ea}{\end{array}}

\newcommand {\eqn}[1]{(\ref{#1})}

\newcommand{\mapru}[1]{\smash{\mathop{\hbox to 1cm{\rightarrowfill}}
\limits^{#1}}}
\newcommand{\maprd}[1]{\smash{\mathop{\hbox to 1cm{\rightarrowfill}}
\limits_{#1}}}
\newcommand{\maplu}[1]{\smash{\mathop{\hbox to 1cm{\leftarrowfill}}
 \limits^{#1}}}
\newcommand{\mapld}[1]{\smash{\mathop{\hbox to 1cm{\leftarrowfill}}
 \limits_{#1}}}
\newcommand{\maprud}[2]{\smash{\mathop{\hbox to 1cm{\rightarrowfill}}
 \limits^{#1}_{#2}}}
\newcommand{\maplud}[2]{\smash{\mathop{\hbox to 1cm{\leftarrowfill}}
 \limits^{#1}_{#2}}}

\newcommand{\lmapru}[1]{\smash{\mathop{\hbox to 1.5cm{\rightarrowfill}}
\limits^{#1}}}
\newcommand{\lmaprd}[1]{\smash{\mathop{\hbox to 1.5cm{\rightarrowfill}}
\limits_{#1}}}
\newcommand{\lmaplu}[1]{\smash{\mathop{\hbox to 1.5cm{\leftarrowfill}}
 \limits^{#1}}}
\newcommand{\lmapld}[1]{\smash{\mathop{\hbox to 1.5cm{\leftarrowfill}}
 \limits_{#1}}}
\newcommand{\lmaprud}[2]{\smash{\mathop{\hbox to 1.5cm{\rightarrowfill}}
 \limits^{#1}_{#2}}}
\newcommand{\lmaplud}[2]{\smash{\mathop{\hbox to 1.5cm{\leftarrowfill}}
 \limits^{#1}_{#2}}}

\newcommand{\cleqn}{\setcounter{equation}{0}}

\begin{abstract}
We study the ``topological gauged WZW model associated with $SU(2)/U(1)$'',
which is defined as the twisted version of the corresponding
supersymmetric gauged WZW model.
It is shown that this model is equivalent to
a topological conformal field theory
characterized by two independent topological conformal algebras,
one of which is the ``twisted Kazama-Suzuki type'' and
the other is ``twisted Coulomb gas type''.  We further show that
our formalism of this gauged WZW model
naturally  reduces to
the well-known  formulations
of 2D gravity coupled with  conformal matter;
one of the gauge choices leads to the K.Li's theory,
and the alternative choices  lead to the KPZ theory or
the DDK (Liouville)  theory.
In appendix we  argue on a possibility of deriving such topological
conformal models
from the $G/G$-gauged WZW models.
\end{abstract}

\newpage

\section{Introduction}
\cleqn

\indent
Conformal models in two-dimensional space-time
possessing $N=2$ world-sheet supersymmetry form a
special class of conformal field theories,
namely they are belived to comprise the only
known solutions to string theory at the perturbative level.

Recently a large class of new $N=2$ superconformal models were constructed
by Y.Kazama and H.Suzuki \cite{KaS}.
They studied coset models  which have $N=1$ superconformal symmetry,
and then determined the precise conditions under which
these $N=1$ models   actually possess $N=2$ supersymmetry.
Some algebraic structures of these models were studied in
\cite{LVW}, \cite{Gepner} emphasizing their geometrical back-grounds.

Subsequently, in the light of the work \cite{TFT},
T.Eguchi and S.Yang
pointed out \cite{EY} that by ``twisting'' the energy-momentum
tensor of these $N=2$ superconformal models with respect to
their $U(1)$-current, they can be interpretted as a kind of topological
conformal field theories with central charge $c=0$.
In the topological conformal models of this type, one of the supercharges
of the original $N=2$ models can be reinterpretted as the BRST charge,
and the chiral primary fields become the physical observables
which are the cohomology classes determined  by this BRST charge.

In general, topological field theories are introduced
by E.Witten \cite{TFT} as the physical theories describing some mathematical
objects, that is, the geometries of various moduli spaces.
It is then interesting to ask if it is possible to describe
the geometrical meanings of the topological conformal models of this type,
in other words the moduli spaces of these models.
Particularly the geometrical interpretation of the BRST cohomology
should be clarified.

{}~

The topological conformal models of this type play
an important role in the recent understanding
of two-dimensional quantum gravity.
Namely the system of topological matter (the topological conformal
model obtained by twisting the $N=2$ minimal model) coupled with
topological gravity \cite{witteng}, \cite{VV}, \cite{KLi} is equivalent
\cite{witten} to the $N$ matrix model \cite{matrix model2}
which is a powerful method for doing non-perturbative
calculations in two-dimensional gravity
\cite{matrix model}.

On the other hand there is still a big gap in the understanding
of the relation between the matrix model formulation and
the conventional approach to two-dimensional gravity based on
conformal field theory coupled to the Liouville theory \cite{Gervais}
(the David-Distler-Kawai (DDK) theory \cite{DDK})
or the Knizhnik-Polyakov-Zamolodchikov (KPZ) theory
(gravity in the light-cone gauge) \cite{KPZ}.
In this sence it is important to clarify the relation between
the topological formulation and other continuum approaches.

{}~

With these motivations we intend to investigate  the topological
conformal models of this type, that is,
obtainable by twisting $N=2$ models, {\em from the standpoint
of two-dimensional Lagrangian field theory\/}.
The reason why we will do is that
our interests are intensively turned to
the global structures of the models, not to the local fluctuations, since
all local degrees of freedom are canceled out in any topological field theory.
The traditional algebraic techniques of conformal field theories,
in which we don't write the Lagrangian explicitely,
are thought to be not so powerful for our purpose.

We begin in section 2 by formulating the ``topological gauged WZW models'',
by twisting the corresponding N=2 supersymmetric
gauged WZW models \cite{Nakatsu} \cite{witten}.
The quantization of the systems is given by the path-integrations of
chiral fields $g$, gauge fields $A$, and two-dimensional metric $\bm{g}$.
The technique of these path-integrations is an application of the formalism
developed by K.Gawedzki, A.Kupiainen \cite{ck}, and D.Karabali, H.J.Schnitzer
\cite{ks}.
Then we specialize to the  case of $SU(2)/U(1)$.
In this section we shall study the path-integration
only for the ``matter part'', that is, the quantization with
the moduli of two-dimensional metric and the gauge fields fixed.
The local operator formalism, which is consistent with the path-integration,
is presented. The matter part is shown to be characterized by two kind
of topological conformal algebras (TCAs).
One of them is a topological conformal algebra obtained by
twisting a $N=2$ superconformal algebra realized by the Kazama-Suzuki
(KS) formalism \cite{KaS}, and the other is
found to be equivalent to that obtained by twisting
the one realized in terms of the ``$N=2$ Coulomb gas formalism'' \cite{CG}.
These topological algebras are not equivalent,
and define respectively the ``Kazama-Suzuki (KS) sector'',
the ``Coulomb gas (CG) sector'' of the matter part.
Investigations of the geometrical meanings of these two sectors are given.

In section 3 we shall discuss the relation with  the well-known
formulation of two-dimensional
gravity coupled to conformal matter.
For this aim, our interests will be poured into
the Coulomb Gas sector rather than the Kazama-Suzuki sector.
There we argue on  the path-integrations of the residual
moduli, i.e. the moduli of the gauge field and the metric.
They lead to the gauge fixed action of the famous formulation
of pure topological gravity by E.Verlinde and H.Verlinde \cite{VV}.
This implies that our model is equivalent to the model considered
by K.Li \cite{KLi}.
By taking alternative gauge fixing procedures,
our model is shown to be equivalent to other continuum approaches
to two-dimensional gravity.
Namely we shall intend to perform
the path-integral over the mode of the Virasoro anomaly,
in place  of the chiral anomaly,
of the matter sector.
This leads to the KPZ theory \cite{KPZ} or the DDK theory \cite{DDK}.

Finally in section 4 we give several discussions.

In appendix A we discuss a possibility of constructing the topological
gauged WZW models from the $G/G$-gauged WZW models, with
a careful restriction on the functional space of the gauge field
over which the path-integral is performed.
Moreover we give some comments on the geometrical back-grounds of
the model.

{}~

\section{Topological Gauged WZW Model associated with $SU(2)/U(1)$}

\cleqn
\subsection{Topological Gauged WZW Models - the definition of the model}

\indent
{}~~~~First of all, we shall present a brief review on gauged WZW models.
Let $(\Sigma , \ca{J})$ be a compact connected Riemann surface
($\ca{J}$ is a complex structure), and  $G$  be a compact semi-simple
Lie group  with its Lie algebra   $\gerg$. Suppose $H$ be a closed
subgroup of $G$ and the corresponding Lie subalgebra $\gerh$.
The action of
the ``$G/H$-gauged WZW model with level $k$'' is defined as follows;
\be
\ba{l}
\dsp
k S_{G} (g,A) =  \frac{ ik}{4 \pi}\int_{\Sigma}
  (g^{-1}\deebar g, g^{-1}\partial g)
  -\frac{ik}{24 \pi}\int_{B}(\tilde{g}^{-1}d\tilde{g},
          \lb \tilde{g}^{-1}d\tilde{g}, \tilde{g}^{-1} d\tilde{g} \rb ) \\
\dsp     ~~~~~ + \frac{ik}{2\pi}
  \int_{\Sigma}\{ -(g^{-1} \deebar g , A^{10}) + (A^{01}, \partial g g^{-1})
       -(A^{01}, Ad(g) A^{10}) + (A^{01}, A^{10}) \} ,
\ea
\label{gwzw action}
\ee
where the chiral field $g$ is $G$-valued and
the gauge field $A= A^{10} + A^{01}$ ($A^{10}$, $A^{01}$
are respectively the holomorphic and anti-holomorphic components of $A$)
is $\gerh$-valued, i.e. $A^{10}$ ($A^{01}$) is $\gerh^{\bc}$-valued
(locally defined) (1,0)-form (resp. (0,1)-form)
satisfying $A^{10 \dag} = - A^{01}$
(``$\dag$'' is the canonical
``hermitian conjugation'' so that $\gerg = \{ u \in \gerg^{\bc}~;~
         u ^{\dag} = -u \}$).
The inner procuct
$(~,~)$ is the Cartan-Killing form  normalized by $(\theta,\theta)=2$
($\theta$ is the highest root of $\gerg$), and $d = \partial + \deebar$
is the canonical splitting defined by $\ca{J}$.
It is well-known that if $\pi_3 (G) = \bz$, that is indeed
our case since $G$ is
compact semi-simple, we must restrict the value of the parameter $k$
in $\bz _{\geq 0}$ \cite{wittenw}.
The non-negativity of $k$ is of course required
in order to define the theory positive-definetely.
This action is manifestly conformally invariant because the  above definition
needs only the complex structure $\ca{J}$, does not need
any metric structure on $\Sigma$.

The most important property of this action is the following
identity (what is called the ``Polyakov-Wiegmann identity'' \cite{pw});
\be
  S_G ({}^{\Om}\! g, {}^{\Om}\! A) = S_G (g,A)- S_G ({\Om}^{\dag} {\Om}, A),
\label{pw identity}
\ee
where we introduce the concept of
``chiral gauge transformation'' (``complex
gauge transformation'') defined for any $H^{\bc}$-valued $\Om$,
\be
\ba{l}
{}^\Om \! g = \Om^{\dag -1}\, g \,  \Om ^{-1}, \\
 ({}^{\Om} \! A)^{10}  =
     Ad (\Om) \,A^{10} - \partial \Om \, \Om ^{-1}.
\ea
\label{cgauge tr rule}
\ee
This identity is nothing but the cocycle condition of  chiral anomaly,
and can be proved by straightforward calculations.
If the chiral gauge transformation $\Om$ is ``unitary'' (i.e. $H$-valued),
this identity \eqn{pw identity} immediately implies the relation
$S_G({}^{\Om}\! g, {}^{\Om} \! A) = S_G (g,A)$.
Because  for any unitary $\Om$, $\Om^{\dag}\Om =1$
holds and this means the 2nd term of
the RHS of \eqn{pw identity} vanishes. In other words
our action $S_G$ has the gauge invariance for the ``vectorial direction''.
But when  $\Om $ is $H^{\bc}/H$-valued
(i.e. an ``axial gauge transformation''), we suffer the chiral anomaly
at the classical level.

The quantization of this model from the standpoint of path-integration is
fully investigated  in \cite{ck} \cite{ks}.
It is shown there that this gauged WZW model describes the so-called
``$G/H$-coset conformal field theory (CFT)''.
Namely the $G/H$-gauged WZW model is one of the solution
of the problem what  Lagrangian field theory describes this coset model.
Our quantization scheme
of the ``topological gauged WZW models'' (defined below)
will be an application
of those works.
It is worthwhile to comment
on the mechanism of appearing the coset CFTs from these models:
The conformal anomaly of the WZW model is originated in its chiral anomaly.
The path-integration of the $H$-gauge field, especially
the integral along the orbits of axial gauge transformations
``absorbes'' the $H$-part of chiral anomaly,
which gives the result; $c_{\msc{gauged WZW}}= c_{G,k} -c_{H,k}$.
This mechanism is very similar as that given in \cite{DDK}.
Namely the Weyl anomaly of the conformal matter is canceled out
by means of the path-integration of the Liouville field.
This is a suggestive point for our later discussions on 2D gravity.

In the papers \cite{Nakatsu}, \cite{witten}
 N=2 supersymmetric extensions of the above gauged WZW models
were presented. Those can be  defined
associated with a general compact K\"{a}hler homogeneous space $G/H$
with $H$ being a closed subgroup of $G$ including
a maximal torus. It is known that the coset space of this type corresponds to
a ``parabolic decomposition'' of $\gerg^{\bc}$;
\bea
\gerg^{\bc} &=& \gerh^{\bc}~ \oplus~ \germ_+~ \oplus~ \germ_-, \nonumber \\
            &=& Z(\gerh^{\bc})~\oplus
                   ~ \gerh_0^{\bc}~ \oplus~ \germ_+~ \oplus~ \germ_- .
\label{parabolic}
\eea
 Here $Z(\gerh^{\bc})$ is the center of $\gerh^{\bc}$,
$\gerh_0^{\bc}$ is the  semi-simple part of $\gerh^{\bc}$.
We have further set
$\dsp \germ_{\pm} =
\sum _{\al \in \bar{\De}_{\pm}} \gerg_{\al} $,
where $\bar{\Delta}_{\pm} =
\Delta(\gerg^{\bc})_{\pm} \setminus \De(\gerh_0^{\bc})_{\pm}$,
$\Delta(\gerg^{\bc})_{\pm}$, $\De(\gerh_0^{\bc})_{\pm}$
are respectively the system of positive (negative) roots of $\gerg^{\bc}$,
$\gerh_0^{\bc}$.
Associated with this palabolic decomposition, we need to  prepare
$\germ_+  \oplus \germ_-$-valued Weyl fermions
$\bm{\psi}$, $\bar{\bm{\psi}}$ (defined with respect
to some spin structure compatible with the complex structure $\ca{J}$)
and a $\gerh $-valued gauge field $ A$, in order to define the desired
supersymmetric model.
So  the ``N=2 supersymmetric gauged WZW model associated with $G/H$''
is given  by;
\be
Z= \int \ca{D}g\ca{D}A\ca{D}\bm{\psi}\ca{D}{\bar{\bm{\psi}}}\,
  \exp \left \lb  - kS_G(g,A) - \frac{1}{ \pi }
\int_{\Sigma} dv(\bm{g}) \,
  \{ (\bm{\psi}, \deebar_{A \bar{z} }\bm{\psi}  )+
(\bar{\bm{\psi}}, \partial_{Az}\bar{\bm{\psi}})\}\right\rb,
\label{susy gwzw}
\ee
where $dv(\bm{g})$ is the canonical volume element defined by
a fixed K\"{a}hler metric $\bm{g}$,
and  $D_A = \partial_A + \deebar_A $ is the canonical splitting of the
 covariant exterior derivative $D_A$.
We can show that this model describes
the N=2 superconformal model (SCF) obtained
by the Kazama-Suzuki supercoset construction for $G/H$ \cite{Nakatsu},
\cite{witten}, \cite{ns 1}.
In particular, in the case that $G/H$ is hermitian symmetric
we can reproduce those given in \cite{KaS}, \cite{LVW}.

However, the models we are wanting now  are topological conformal models,
namely the ``twisted'' version \cite{EY} of these N=2 SCFs.
{\em Therefore we shall
regard the fermionic fields $\bm{\psi}$, $\bar{\bm{\psi}}$
as a BRST ghost system. That is to say, we shall replace
 the $\germ_+$, $\germ_-$-components
of $\bm{\psi}$
 by a $\germ_+$-valued $(0,0)$-form $\psi$ (ghost),
a $\germ_-$-valued $(1,0)$-form $\chi$ (anti-ghost)
respectively and similarly the $\germ_-$, $\germ_+$-components of
$\bar{\bm{\psi}}$  by a
$\germ_-$-valued $(0,0)$-form $\bar{\psi}$,
$\germ_+$-valued $(0,1)$-form $\bar{\chi}$.}
We shall call the supersymmetric gauged WZW model \eqn{susy gwzw}
with this ``twist'' of the fermionic fields as
the ``topological gauged WZW model associated with $G/H$'',
or briefly the ``topological $G/H$-model''.
The following part of this paper will be devoted to the most simple case;
$G=SU(2)$, $H=U(1)$. Full investigations of the above models
for  general homogeneous spaces
$G/H$, along the scheme of path-integral quantization  performed
in the following sections,
will be presented  in \cite{ns 1}, \cite{ns 2}
on both the untwisted (N=2 SCF)
and the twisted (topological) version.
Especially it can be  shown \cite{ns 1} that the topological $G/H$-model
indeed gives the twisted Kazama-Suzuki model for $G/H$.

Another derivation of the topological gauged WZW model from
the $G/G$-gauged WZW model (manifestly a theory of $c=0$)
is speculated  in appendix A,
which needs a careful restriction of the domain
over which the path-integration is performed in order to get non-trivial
(finite) physical degrees of freedom.

{}~


\subsection{Quantization of the Topological $SU(2)/U(1)$-Model}

\indent
{}~~~~Now let us begin
investigations of the topological $SU(2)/U(1)$-model,
i.e. $G=SU(2)$, $H=U(1)$,
$\gerg =\ger{s}\ger{u}(2)$, $\gerh =\ger{u}(1)$.
Let
\be
\gerg^{\bc} = \gerh^{\bc} \oplus \gerg_+ \oplus \gerg_-
\ee
be the usual Cartan decomposition,
which is  the parabolic decomposition \eqn{parabolic} in this case.
Therefore,
the ghost fields $\psi$, $\bar{\psi}$ should  be
$\gerg_+$, $\gerg_-$-valued (0,0)-forms,
and the anti-ghost fields  $\chi$, $\bar{\chi}$ are
$\gerg_-$-valued (1,0)-form, $\gerg_+$-valued (0,1)-form respectively.
The gauge field $A$ should be $U(1)$ (Cartan) valued.
Let us introduce the standard canonical Cartan-Weyl base
$e$, $f$, $t$ of $\gerg^{\bc}$;
\be
e = \left(
\ba{rr} 0& 1 \\
        0& 0 \ea \right),~~~
f= \left(
\ba{rr} 0& 0 \\
        1& 0 \ea \right),~~~
t= \left(
\ba{rr} 1& 0 \\
        0& -1 \ea \right).
\ee
They satisfy the relations;
\be
\ba{l}
(t,t)= 2,~~~(e,f)=1, \\
\mbox{other combinations vanish},
\ea
\ee
with the usual definition of the Cartan-Killing metric for $\gerg^{\bc}$;
\be
(A,B) \equiv \mbox{Tr}(AB).
\ee
In terms of  this base, let us write  the ghost system
and the gauge field introduced above by components;
$\psi e$, $\bar{\psi} f$, $\chi f$, $\bar{\chi} e$,
and $\dsp A \frac{i}{2}t$ (so that $A$ becomes  a real 1-form on $\Sigma$).
Then the model we bigin with is defined by\footnote
  {In this paper we shall take the convention that   fermionic fields
   and  fermionic (BRST) transformations should {\em anti-commute\/}
   with 1-forms on the space-time $\Sigma$ (i.e. $dz$, $d\bar{z}$).
   The motivation of it is due to the following fact; In field
   theory, BRST ghost fields
   should be identified with differential forms
 on some functional space, and particularly in topological
   field theory it is often convenient to consider the product space
  of the space-time and some moduli space.};
\bea
Z  & =& \int  \ca{D}\bm{g}\ca{D}g \ca{D}A
   \ca{D}(\chi, \bar{\chi}, \psi, \bar{\psi} )
     \, \exp \left\lb -kS_G(g,A) - \frac{1}{2 \pi i} \int_{\Sigma}
  (\deebar_{A}\psi\, \chi -\bar{\chi}\, \partial_{A}\bar{\psi})
   \right\rb
\label{gwzw model} \\
&\equiv & \int \ca{D}\bm{g}\, Z\lb \bm{g} \rb ,  \nonumber\\
\eea
where $\bm{g}$ denotes the metric on $\Sigma$.
Although the gauged WZW action $S_G (g, A)$ and the ghost's action
does not depend on the metric $\bm{g}$ (since they depend only on $\ca{J}$),
the path-integral measures $\ca{D}A$, $\ca{D}(\chi, \bar{\chi},
\psi, \bar{\psi})$ implicitely depend on $\bm{g}$, which appears as
the Virasoro anomalies (the Weyl or the gravitational anomalies).
In the following,
we suppose that {\em the metric $\bm{g}$ defines the complex structure
$\ca{J}$ so that $\bm{g}$ is a K\"{a}hler metric
with respect to $\ca{J}$.}
The covariant exterior derivatives are explicitely written as;
\be
D_A \psi = d \psi + i A \psi, ~~~ D_A \bar{\psi}
= d \bar{\psi} -i A \bar{\psi},\ee

Our fundamental standpoint of quantization
is that we shall perform the path-integration
of {\em all\/} fields appearing in the theory, i.e. the chiral field $g$,
the gauge field $A$, the ghost fields $\chi$, $\bar{\chi}$, $\psi$,
$\bar{\psi}$
and the metric $\bm{g}$.
However,
in this section we shall  perform the path-integration
only for the ``matter part'', that is, quantize this model with the moduli
of the metric and the gauge field  fixed.
On these integrals of the moduli
we  will  discuss in the next section.

The action \eqn{gwzw model} has a  following  ``on-shell'' BRST symmetry,
which is originated in the N=2 SUSY of the untwisted model;
\be
\ba{l}
\delta_{G/H} \chi = k(e,\partial_A g\, g^{-1} ),~~~
   \bar{\delta}_{G/H} \bar{\chi}
     = - k (f, g^{-1}\deebar_A g),\\
\delta_{G/H}g = (\psi e) g,~~~
  \bar{\delta}_{G/H}g = - g (f\bar{\psi}) ,\\
\mbox{other combinations are defined to vanish.}
\ea
\label{susy brst}
\ee
This action further has the $U(1)$-gauge symmetry (along the vectorial
direction),
but suffer the chiral anomaly for axial gauge transformations,
in the similar manner as the usual $SU(2)/U(1)$-gauged WZW model.
We must perform the  gauge fixing  for these  $H^{\bc}$-chiral
gauge transformations,
{\em not\/} only for  vectorial gauge transformations.
It may be  somewhat confusing to call it ``gauge fixing'',
since the chiral anomaly exists.
More properly we should call it a  transformation of the
path-integration variables,
namely from the measure of $U(1)$-gauge field to the product
of the measure along the chiral gauge orbits and the measure of the modulus.
According to the Fadeev-Popov (FP) like prescription,
we shall insert the following identity into \eqn{gwzw model};
\be
1 = \int \ca{D}a \ca{D}X \ca{D}Y
\delta (A- {}^h \! a)
      \De_{\msc{FP}}(a)
\ee
where $\dsp h = e^{X+ iY}$
is a chiral gauge transformation
for $H^{\bc}$ ($X$, $Y$ are real scalar fields,
and now we choose $\dsp \frac{1}{2}t$ as the $U(1)$-generator),
$a$ is the ``back-ground gauge field''
which describes the modulus of
the $U(1)$-gauge field $A$, and $\dsp \int \ca{D}a$ is
in effect nothing but
a finite-dimensional integral.
``$\De_{\msc{FP}}(a)$'' is a well-known FP determinant (
the Jacobian between the
path-integral measures $\ca{D}({}^h \! a)$ and $\ca{D}X\ca{D}Y$),
which can be rewritten
by additional FP ghosts;
$\dsp \xi \frac{1}{\sqrt{2}}t$ ($\gerh^{\bc}$-valued (0,0)-form, ghost),
$\dsp \bar{\xi} \frac{1}{\sqrt{2}}t$ ($\gerh^{\bc}$-valued (0,0)-form, ghost),
$\dsp \zeta \frac{1}{\sqrt{2}}t$ ($\gerh^{\bc}$-valued (1,0)-form,
anti-ghost),
$\dsp \bar{\zeta} \frac{1}{\sqrt{2}}t$ ($\gerh^{\bc}$-valued (0,1)-form,
anti-ghost),
\be
\De_{\msc{FP}}(a)= \int \ca{D}(\zeta, \bar{\zeta}, \xi, \bar{\xi})
   \, \exp \left\lb - \frac{1}{2 \pi i} \int_{\Sigma}  (\deebar  \xi \, \zeta
-\bar{\zeta}\, \partial \bar{\xi})    \right\rb.
\ee
Notice that $\De_{\msc{FP}}$ is indeed independent of the back-ground
gauge field $a$, since $H^{\bc}$ is abelian.

We should remark that the integral
$\dsp \int \ca{D}Y$ (the vectorial direction)
is a gauge volume,
but $\dsp \int \ca{D}X$ (the axial direction) must be performed
because of the existence of  chiral anomaly.
We need to estimate the chiral anomalies for the chiral field $g$ and
the ghost system $\chi$, $\bar{\chi}$, $\psi$, $\bar{\psi}$ independently.

For $g$, by the Polyakov-Wiegmann identity \eqn{pw identity}, we obtain
\be
\ba{lll}
S_G (g, {}^h\! a) &=& S_G ({}^{h^{-1}}\! g, a) - S_G(h^{\dag}h,a) \\
 \dsp   & = &  \dsp S_G ({}^{h^{-1}}\! g, a)
   - \frac{i}{2\pi} \int_{\Sigma} \{ \deebar X \partial X + i X F(a) \}
\ea
\ee
where $F(a) = da$ is the curvature of the back-ground gauge field $a$.
Moreover since the measure $\ca{D} g$ has no anomaly,
it holds that
\be
\int \ca{D}g \, \exp \{ -k S_G ({}^{h^{-1}}\! g, a) \}
= \int \ca{D}g \, \exp\{  -k S_G ( g, a) \} .
\ee

On the other hand, for the ghost system $\chi$, $\bar{\chi}$, $\psi$,
$\bar{\psi}$;
\bea
Z_{\chi \psi} &= &
    \int \ca{D}(\chi,\bar{\chi}, \psi, \bar{\psi})\,
    \exp \left\lb -\frac{1}{2\pi i} \int_{\Sigma}
    (\deebar_{{}^h \! a}\psi \, \chi \, -\bar{\chi}\, \partial_{{}^h \! a}
      \bar{\psi})   \right\rb \\
    &=& \int \ca{D}(\chi,\bar{\chi}, \psi, \bar{\psi})\,
    \exp \left\lb -\frac{1}{2\pi i} \int_{\Sigma}
    (\deebar_{a}\psi\, \chi \,
     \, -\bar{\chi}\, \partial_{a}\bar{\psi}) \right\rb
    \nonumber \\
    & & ~~~\times \exp \left\lb \frac{ i}{ \pi}
      \int_{\Sigma}  \{ \deebar X\partial X
     +iX F(a)  + \frac{1}{2 i}X R(\bm{g})\} \right\rb ,
\label{determinant chi psi}
\eea
where
$\scs{  \left(
     \ba{cc}
       0    & R(\bm{g}) \\
       -R(\bm{g}) & 0
     \ea
      \right) } $
is the curvature of the Levi-Civita connection
$\scs  {\left( \ba{cc}
      0   &  \om(\bm{g})  \\
    -\om(\bm{g})& 0   \ea \right) } $
with respect to $\bm{g}$.
This abelian anomaly appears as the
``back-ground charge''. It can be easily estimated by  direct
computations or more elegantly by making use of the index theorem
associated with  the twisted Dolbeault complex (refer for example \cite{EGH});
$O~ \mapru{~}~ \Om^{(0,0)}~ \mapru{\deebar_a}~ \Om^{(0,1)}~ \mapru{~}~ O$
with respect to the ghost system $\chi$, $\psi$.
This leads that the back-ground charge of the ghost's gauge current
(which is nothing but the  ``$-$ ghost number current'') is equal to
$\dsp \frac{1}{2} \chi (\Sigma) + C_1$.
Here $\chi(\Sigma)$ stands for  the
Euler number of $\Sigma$ and $C_1$ means the 1st Chern number of
the holomorphic line bundle  such that
$\psi$ is one of the section of it.
The terms $\dsp \frac{1}{2}\chi(\Sigma)$, $C_1$ correspond
respectively to the terms
$\dsp \frac{1}{2\pi} \int_{\Sigma} X R(\bm{g})$,
$ \dsp -\frac{1 }{ \pi} \int_{\Sigma} X F(a)$ in \eqn{determinant chi psi}.

Taking all things into account,
we can arrive at the gauge fixed model;
\bea
 Z_{\msc{gauge}} \lb \bm{g} \rb &=&
   \int \ca{D}a\, Z_{\msc{gauge}} \lb \bm{g}, a \rb  \nonumber \\
Z_{\msc{gauge}} \lb \bm{g}, a \rb & = &
  \int   \ca{D}(g, X,  \chi,\bar{\chi},\psi,\bar{\psi},
     \zeta, \bar{\zeta}, \xi, \bar{\xi} )  \nonumber \\
  &  &  ~~\times \exp \{ -k S_G ( g, a )  -
 S_{\chi \psi} (\chi,\bar{\chi}, \psi,\bar{\psi}, a) \}
\label{2nd step gauge fixing} \\
&  &  ~~\times \exp \{ - S_X (X , a, \om(\bm{g}))
      - S_{\zeta \xi} (\zeta, \bar{\zeta}, \xi, \bar{\xi} ) \}. \nonumber
\eea
In \eqn{2nd step gauge fixing} we introduce  the following notations;
\bea
S_{\chi \psi} (\chi,\bar{\chi}, \psi,\bar{\psi}, a )
& = & \frac{1}{2\pi i} \int_{\Sigma}
    (\deebar_{a}\psi\, \chi \, -\bar{\chi}\, \partial_{a}\bar{\psi}),
      \label{action chi psi} \\
S_{\zeta \xi} (\zeta, \bar{\zeta}, \xi, \bar{\xi}  )
& = & \frac{1}{2 \pi i} \int_{\Sigma} (\deebar  \xi\, \zeta
\, -\bar{\zeta}\, \partial \bar{\xi})    , \label{action zeta xi}\\
S_X (X , a, \om(\bm{g}))
& = & \frac{ 1}{2 \pi i} \int_{\Sigma}\{  \deebar X \partial X
     + i \al_+  X F(a)
     + i \al_- X R(\bm{g}) \} , \label{action X}\\
& & ~~~ (\al_+ = \sqrt{k+2},
      ~ \al_- = - \frac{1}{\sqrt{k+2}}),  \nonumber
\eea
where we have properly rescaled the scalar field $X$ in \eqn{action X}.

{}~

Let us turn to the local operator formalism.
We fix a coordinate neighborhood $U \subset \Sigma$ and a holomorphic
coordinate system $z$ on $U$,
in which we work. For the time being we shall set $a=0$, $\om (\bm{g})=0$
on $U$ in order to make things easy, which is of course always possible if
$U$ is sufficiently small.

{}From the gauge fixed action
\eqn{2nd step gauge fixing}
we can immediately compute
the total energy-momentum (EM) tensor $T_{\msc{tot}}$ of the matter part
(we treat only the holomorphic sector);
\be
T_{\msc{tot}} =  T_g + T_X + T_{\chi \psi} + T_{\zeta \xi}.
\label{ttotal}
\ee
In \eqn{ttotal} $T_g$ is the EM tensor of the $SU(2)$-WZW model
(the Sugawara EM tensor) and $T_X$ is that of the real scalar field $X$.
$T_{\chi \psi}$, $T_{\zeta \xi}$
are those of the  ghost fields. Their explicit forms are given by;
\bea
T_g &= & \frac{1}{2(k+2)}:(J_g, J_g):,  \\
T_X & = & -:(\partial_z X)^2: + \al_- \partial_z^2 X, \\
T_{\chi \psi} & = & - :\chi_z\, \partial_z \psi:,~~~~
T_{\zeta \xi} = - :\zeta_z\, \partial_z \xi :,
\eea
where $J_g = -k\partial_z g \, g^{-1}$ is the chiral current of the
$SU(2)$-WZW model, and
``:~~:'' stands for the usual normal ordering.
The free scalar field $X$ and the ghost fields
$\chi$, $\psi$, $\zeta$, $\xi$ satisfy  the following
operator product expansions
(OPEs);
\be
\ba{l}
\dsp \partial_z X(z) \partial_w X(w) \sim -\frac{1}{2(z-w)^2} , \\
\dsp \chi_z (z) \psi (w) \sim \frac{1}{z-w}, ~~~
\zeta_z(z) \xi(w) \sim \frac{1}{z-w}.
\ea
\ee
Moreover
we express the chiral current
$J_g$ by components for later convenience;
\be
J_g = J^0_g t + J^+_g f + J^-_g e,
\ee
with the following OPEs satisfied;
\be
\ba{l}
\dsp
J^0_g (z)J_g^{\pm}(w) \sim \frac{\pm 1}{z-w}J_g^{\pm}(w),
{}~~~ J^0_g (z) J^0_g (w) \sim \frac{k}{2(z-w)^2},   \\
\dsp
J^+_g (z) J^-_g (w) \sim \frac{k}{(z-w)^2} + \frac{2}{z-w}J^0_g (w), \\
\mbox{other combinations have no singular OPEs.}
\ea
\ee
{}From these explicit forms it is easy to show that
$T_{\msc{tot}}$ has vanishing central charge;
\be
\ba{lll}
c_{\msc{tot}} &=& c_g + c_X + c_{\chi\psi}+c_{\zeta\xi} \\
             & =& \dsp \frac{3k}{k+2} + (1 + 6 \al_-^2) + (-2) +(-2)=0.
\ea
\ee
where $c_g$, $c_X$, $c_{\chi \psi}$, $c_{\zeta \xi}$ are
the central charges of
$T_g$, $T_X$, $T_{\chi \psi}$, $T_{\zeta \xi}$ respectively.
Thus the matter part of our model is a topological
conformal model   as was expected.

In order to extract the physical degrees of freedom
from our total matter system,
it is important to construct the BRST complex
of which cohomology classes describe them.
{}From our treatment of the gauge symmetry our BRST complex
should be characterized by two kind of BRST charges.

The  on-shell  BRST symmetry
(supersymmetry) \eqn{susy brst}
should be  characterized by the following BRST charge;
\be
Q_{G/H} =   \frac{1}{2\pi i} \oint dz \, G^+_{G/H}  ,
\ee
where the BRST current $G^+_{G/H}$ is defined as
\be
G^+_{G/H} =  -\al_-  \psi J^+_g.
\label{g+gh}
\ee
Similarly the  gauge fixing for the $H^{\bc}$-gauge transformations
 should be  performed by
\bea
Q_{H^{\bc}} &=&  - \frac{\al_-}{2\pi i}
\oint dz \, \xi J^0_{\msc{tot}}  \label{qhc0}\\
   & = & \frac{1}{2\pi i} \oint dz \, G^+_{H^{\bc}} ,
\eea
where we define
\be
G^+_{H^{\bc}} =
      -\al_-  (\xi J^0_{\msc{tot}}
      - \partial_z \xi) .
\label{g+hc}
\ee
(Of course the term ``$\partial_z \xi$'' does not contribute,
since it is a total derivative.)
In \eqn{qhc0}, \eqn{g+hc} $J^0_{\msc{tot}}$ is the ``total $U(1)$-current''
of the model;
\be
J^0_{\msc{tot}} = J^0_g + J^0_{\chi \psi} + J_X^0
  \equiv \hat{J}^0 + J_X^0,
\ee
where
$\hat{J}^0 = J_g^0 + J_{\chi \psi}^0$
is the $U(1)$ current associated with the $U(1)$-gauge symmetry,
and $J_X^0$ is that of
the gauge field. Their explicit forms are easily computed from
\eqn{action chi psi}, \eqn{action X};
\bea
J_{\chi \psi}^0 & =& :  \chi_z  \psi  :,\\
J_X^0 & = & \al_+ \partial_z X  .
\eea

Now it may be useful to give some comments on the $U(1)$-currents
appeared above.
The $U(1)$-currents $\hat{J}^0$, $J^0_{\msc{tot}}$ are invariant
under the $Q_{G/H}$-transformation, and are not exact with respect to
this BRST charge. Especially the total $U(1)$-current $J^0_{\msc{tot}}$
has no Schwinger term
in its current-current OPE.
This guarantees the nilpotency of  the BRST charge
$Q_{H^{\bc}}$. (c.f. \cite{ks})
It is of course $Q_{H^{\bc}}$-trivial;
\be
J^0_{\msc{tot}} = \{ Q_{H^{\bc}},\,  \al_+ \zeta_z  \},
\label{jhtot brst}
\ee
as is a common feature of the standard BRST theory.

We can also  check
that  $Q_{G/H}$ and $Q_{H^{\bc}}$ anti-commute with each other.
Hence our  BRST complex,
whose differential is given by  the total BRST charge
$Q_{\msc{tot}} = Q_{G/H} + Q_{H^{\bc}}$,
has a double complex structure.
The physical observables are constructed as the cohomology classes
of this double complex.

In order to have some insight into the above cohomological
problem we give attention to the fact that $T_{\msc{tot}}$
can be written as the following BRST exact forms;
\bea
T_{\msc{tot}} & =& \{ Q_{\msc{tot}},\,
             G^-_{G/H}+G^-_{H^{\bc}} \}  \\
&=& \{ Q_{G/H}, G^-_{G/H} \} + \{ Q_{H^{\bc}}, G^-_{H^{\bc}}\} \nonumber \\
& =& T_{G/H} + T_{H^{\bc}}, \nonumber
\eea
where $G^-_{G/H}$, $G^-_{H^{\bc}}$ are defined as
\bea
G^-_{G/H} &= & -\al_-  \chi_z  J_g^- , \label{g-gh}\\
G^-_{H^{\bc}} &= &
   -\al_-  \{ \zeta_z  (\hat{J}^0 - J^0_X)
     -\partial_z \zeta_z \}, \label{g-hc}
\eea
and we set
\bea
 T_{G/H}& = & \{ Q_{G/H},  \, G^-_{G/H}\},   \nonumber \\
         & = &  \frac{1}{2(k+2)} \{ :(J_g, J_g): -
  :2 (\hat{J}^0)^2  : \}  \label{tgh} \\
  & & ~~~~ + \frac{1}{k+2} \partial_z \hat{J}^0
       - :\chi_z \, \partial_z \psi :,   \nonumber \\
 T_{H^{\bc}}& =& \{ Q_{H^{\bc}}, \,  G^-_{H^{\bc}}\} \nonumber \\
       & = & \frac{1}{(k+2)} :(\hat{J}^0)^2:
      - \frac{1}{k+2} \partial_z \hat{J}^0 \label{thc} \\
  & & ~~~~ - :(\partial_z X)^2  : + \al_- \partial_z^2 X
     - :\zeta_z  \partial_z \xi: . \nonumber
\eea
{}From the look of \eqn{g+gh}, \eqn{g-gh}, and \eqn{tgh} it is clear that
{\em $\{  T_{G/H}, ~ J_{\msc{KS}},~ G^+_{G/H},~ G^-_{G/H}\} $
generate the topological conformal algebra (TCA)
constructed by twisting the Kazama-Suzuki model
for $SU(2)/U(1) =\bc P^1$ $\dsp (c=c_k\equiv \frac{3k}{k+2})$ \cite{KaS},
with the familiar definition of  $U(1)$-current};
\be
J_{\msc{KS}} = :\psi  \chi_z: +
\frac{2}{k+2} \hat{J}^0 .
\ee
Here we notice that the generators $T_{G/H}$, $J_{\msc{KS}}$,
and $G^{\pm}_{G/H}$ are invariant under the $Q_{H^{\bc}}$-transformations.
We shall call this TCA as
the ``Kazama-Suzuki (KS) sector''
of the  topological $SU(2)/U(1)$-model.

On the other hand,
suppose we bosonize the $U(1)$-current $\hat{J}^0$
by introducing a real scalar boson $\vphi$
compactified in the circle with the radius $\al_+$
(normalized by
$\dsp \partial_z \vphi (z)
\partial_w \vphi (w) \sim  -\frac{1}{2(z-w)^2}$);
\be
 \hat{J}^0  \equiv i \al_+ \partial_z  \vphi ,
\ee
and further define a  complex boson $\hat{\vphi}$ as
$\hat{\vphi} = \vphi - iX$.
We can easily obtain the following expressions for $T_{H^{\bc}}$
\eqn{thc},
$G_{H^{\bc}}^+$ \eqn{g+hc} and $G_{H^{\bc}}^-$ \eqn{g-hc}
in terms of $\hat{\vphi}$;
\bea
T_{H^{\bc}} &= &-: \partial_z \hat{\vphi}^{\dag} \partial_z \hat{\vphi} :
 + i \al_- \partial_z^2 \hat{\vphi}
              - :\zeta_z \partial_z \xi:,       \label{thc'}\\
G^+_{H^{\bc}} &= &  i \xi \partial_z \hat{\vphi}
     +\al_- \partial_z \xi  ,\label{g+hc2} \\
G^-_{H^{\bc}} &= &  i \zeta_z \partial_z \hat{\vphi}^{\dag}
     +\al_- \partial_z \zeta_z , \label{g-hc2}
\eea
(Notice the relation
$J^0_{\msc{tot}} = i \al_+ \partial_z \hat{\vphi}$.)

{\em Surprisingly this EM tensor \eqn{thc'},  the fermionic currents
$ G_{H^{\bc}}^+$ \eqn{g+hc2},
$ G_{H^{\bc}}^- $ \eqn{g-hc2},
and the  $U(1)$-current};
\be
J_{\msc{CG}} = :\xi \zeta_z :
+ i\al_- \partial_z \hat{\vphi} -  i\al_- \partial_z \hat{\vphi}^{\dag}
\label{jcg}
\ee
{\em generate another TCA,
which precisely coincides with that constructed
by twisting the so-called ``Coulomb gas representation'' of the $N=2$
minimal model \cite{CG}.
That is to say, they can be identified
with that of the topological matter system in the K.Li's
theory of 2D gravity!! \cite{KLi}}
We shall call this TCA  the ``Coulomb gas (CG) sector''
in contrast to the Kazama-Suzuki sector.

These two TCAs are completely isomorphic to each other
because their untwisted $N=2$ superconformal algebras are both of
$\dsp c= \frac{3k}{k+2}$ (i.e. the minimal model).
But it should be noticed that
they are ``essentialy'' independent.
To be specific, $G^+_{G/H}$, $G^-_{G/H}$,  $T_{G/H}$,
$\dsp q_{\msc{KS}}\equiv \frac{1}{2\pi i} \oint dz \, J_{\msc{KS}}$
(not $J_{\msc{KS}}$ itself!)
commute (or anti-commute for fermionic currents) with
 the corresponding operators of the CG sector
and they are not BRST-equivalent.
Therefore these two independent TCAs
should correspond to  different geometrical degrees of freedom,
that is, describe different moduli spaces.

However, as will be seen below,
in the set of  physical observables of the total  system
we can find  the objects naturally supposed to be common objects of
the two sectors, which are no other than the ``chiral primary fields''.
This aspect is one of the interesting problems in  studies of the
topological gauged WZW models.
The similar phenomena in the higher rank cases
will be  seen in \cite{ns 1}.

Let us study
the physical observables of the model, anticipating the appearance of
two kinds of moduli spaces.
Of course they are  defined as the BRST cohomology classes
with respect to the total
BRST charge $Q_{\msc{tot}}$.
If degeneration occurs at the ``$E_2$ term'' of the spectral sequence
defined by this total double complex,
one can precisely compute the total cohomology,
by taking cohomology first by $Q_{G/H}$ and then by $Q_{H^{\bc}}$
or vice versa.
We can see that, when taking cohomology first by $Q_{G/H}$,
the system is characterized by the CG sector, while,
when taking cohomology first by $Q_{H^{\bc}}$, it is characterized
by the KS sector.
Assuming the above degeneration, we may conclude that
there exists a kind of ``duality'' between those two sectors.
Related with this observation it is appropriate to note that
 the following simple relation between
$J_{\msc{KS}}$ and $J_{\msc{CG}}$ holds;
\be
J_{\msc{KS}} - :\psi \chi_z: = J_{\msc{CG}}- :\xi \zeta_z: ~~~
(\mbox{modulo BRST} ).
\ee
This suggests that the bosonic parts of the physical observables
for  the two sectors are completely common objects. The  difference
between them consists in the ghost sector.

Although we have not yet succeeded in the
complete resolution of this total BRST complex,
we can present the following set of  physical observables;
\be
\ca{O}_{k,l,n} (z) = \,
  :\psi(z)^k \, e^{-ik\al_- \hat{\vphi}^{\dag}(z)} \,
\xi(z)^l \, e^{i n\al_- \hat{\vphi}(z) } :,
\label{physical observables}
\ee
where $k,l=0,1$,  $n \in \bz$.
Here  we omitted the ``super partners'',
which can be constructed
from \eqn{physical observables}
by making use of $G^-_{G/H}$, $G^-_{H^{\bc}}$.
It should be remarked that the scalar field $\hat{\vphi}$ and the ghost
$\psi$ are not independent;
\be
 i\al_+ \partial_z \hat{\vphi} (z) \psi(w) \sim
                 - \frac{1}{z-w} \psi(w),
\ee
thus we need the term $e^{-ik\al_- \hat{\vphi}^{\dag}(z)}$
in order to restore the BRST invariance.

At  first glance, it seems that there exist infinite dimensional
physical degrees of freedom. However, more strictly we must further
restrict the physical Hilbert space,
since the Hilbert space characterizing the theory should be
constructed from that of the original $SU(2)$ WZW model,
which is smaller than the Fock space of $\vphi$.
This  leaves
at most finite physical degrees of freedom,
because the WZW model with level $k\in \bz_{\geq 0}$
possesses only finite primary fields.
Considering the no-ghost sector, we can  take the following
subset of \eqn{physical observables};
\be
\ca{O}_j (z)  = \, :e^{-i j \al_- \hat{\vphi}(z)} :,
{}~~~(j = 0, 1, \cdots , k).
\ee
These are precisely the same as the ``chiral primary fields''
(of the ``$A_{k+1}$-type'') in the untwisted $N=2$ minimal model.
It is worth remarking that these chiral primary fields correspond to
the primary states having the forms;
$| j,j \rangle _g \otimes | -j \rangle_X $ $(j=0,1,\cdots, k)$,
where ``$| j,j \rangle _g $'' means the ``highest weight component of
the primary states with spin $j/2$'' of $g$ and ``$| -j \rangle_X$''
stands for the primary state of $X$ having a value $-j$ as the
$U(1)$-charge for $J^{\gerh}_X$.

Let us argue on the geometrical interpretation of the model.
The moduli space (of the matter system)
is thought to be  a K\"{a}hler space;
\be
G/H \times H^{\bc} \cong \bc P^1 \times
\left( \bc  P^1-\{ 0, \infty \} \right),
\ee
(roughly speaking, the ``target space'' of our gauged WZW model.)
The coset space $G/H$ corresponds to the KS sector, and $H^{\bc}$
corresponds to the CG sector.
It is clear from the construction that the moduli space for the CG sector
is indeed so, since it should correspond to the  constant modes of chiral
gauge transformations.
However, as regards the KS sector it is not so manifest what
  geometrical meanings are included
from our starting point, i.e. the twisted N=2 supersymmetric gauged WZW
model. In fact the BRST transformations of the KS sector are the
twisted version of the supersymmetric transformations, so  naively
it is unclear what meanings its BRST cohomology should have.
But, as was already commented,
 our topological gauged WZW action \eqn{gwzw model}
can  be also  derived from  the $G/G$-gauged WZW model
(see appendix), in which
the BRST transformations for the KS sector \eqn{susy brst}
is  obtained by performing the gauge fixing corresponding
to the gauge degrees of freedom of the $G/H$-part
of  gauge transformations.
Therefore we can regard its moduli space as above, and further
give the following simple geometrical observation:
The zero-modes of $\psi$, $\bar{\psi}$
(the ghosts for the KS sector)  are
identified with  $(1,0)$-  and $(0,1)$-forms
on the moduli space $G/H$,
 the BRST charges $Q_{G/H}$, $\bar{Q}_{G/H}$ are respectively
identified with the holomorphic and anti-holomorphic parts of the exterior
derivative on  this K\"{a}hler space.
The similar observation is possible  with respect to
the $H^{\bc}$-part of the moduli space, the ghosts for the CG sector
$\xi $, $\bar{\xi}$, and the BRST charges $Q_{H^{\bc}}$,
$\bar{Q}_{H^{\bc}}$.

It is a challenging problem how  we should interpret
geometrically the bosonic part
(i.e. the $\hat{\vphi}$-part) of the chiral primary fields.
Remarkably they have fractional $U(1)$-charges, and so it is difficult
to interpret them as  differntial forms on some moduli spaces
like as ghost fields.
In this respect it may be worthwhile to keep the following fact in mind:
Our  Lagrangian defining the model \eqn{gwzw action}
is  neither zero (nor BRST exact) nor some topological invariants,
namely our theory is {\em not\/} topologically invariant
{\em at the classical level.} This is the main distinction
of our model
from the topological field theories which  start with
zero Lagrangians (for example see \cite{TFT2}, \cite{witteng}, \cite{VV}).
In these theories  all the quantum effects are canceled out,
and  all the  concepts
can be translated into mathematical languages.
In particular
the objects with fractional $U(1)$-charges
as above do not come into sight.
But our model is a theory with an essentially non-zero Lagrangian.
It is thought that {\em the quantum effects are not completely
canceled out, although we have still only topological invariants as the
physical observables.} We might be able to say
that the appearance of these objects is a purely quantum effect.

On this problem, the formulation presented in \cite{witten} is interesting
and suggestive from geometrical  viewpoints.
The author introduces the concept such as the ``$(k+2)$th root of
the canonical line bundle over the Riemann surface'' in order to deal with
the above objects with fractional $U(1)$-charges in the topological matter.
The difference between the formulation given in \cite{witten} and
 ours mainly consists in the choice of the gauge fixing prescriptions:
In \cite{witten} the gauge fixing is performed by a kind of ``fixed point
manipulation'', in which the physical degrees of freedom
are directly extracted by some simple geometrical argument,
and then the model reduces to ``the abelian model'',
that is, the topological $U(1)$ gauge theory coupled to 2D gravity.
Meanwhile our prescription given above is the standard BRST-procedure,
in which we design to confine the unphysical (gauge) degrees of
freedom by introducing the new unphysical ones, that is, the
(non-zero modes of) ghost fields.
It may be plausible that the ``fixed point manipulation'' in \cite{witten}
corresponds to the Lefschetz formula applied to the BRST complex
with $Q_{G/H}$, and that ``the abelian model'' in \cite{witten}
is related with the CG sector obtained above \cite{ns l}.

Taking account of this consideration
and the fact that the twisted Coulomb-gas model
is no other than the K.Li's minimal topological matter \cite{KLi},
we may well say that the CG sector will play an important role
in order to specify the relation with 2D gravity.
For this purpose we think it
convenient to drop the KS sector out in our model.

{}~

\section{Relation with 2D Gravity}

\cleqn

\subsection{Relation with the K.Li's Theory of 2D Gravity}

\indent
{}~~~~In the previous section we have found out that the topological
$SU(2)/U(1)$-model is characterized by  two independent
topological conformal algebras; the KS sector (``the coset part'')
and the CG sector (``the Cartan part'').
In this section we would like to discuss the coupling to 2D gravity
of our model.
As was already mentioned,
the CG sector is thought more important than the KS sector
for this aim.
Therefore, in the following arguments of this paper
we will kill the KS sector out.
This only means
to  drop the zero-modes of the ghosts $\chi$, $\psi$ by taking the
equivariant cohomology in defining the physical states,
since the non zero-modes of the ghost fields are already suffered
quartet confinements \cite{ns l}.

To complete the quantization of our model, we must integrate the residual
 degrees of freedom, i.e. the back-ground gauge field $a$ and
the metric $\bm{g}$.
Introducing a zweibein $(e^+, e^-)$ $(\overline{(e^+)} = e^-)$
instead of $\bm{g}$ and then defining an
$ISO(2)$-gauge field $\ca{A} = (a, e^+,e^-)$\footnote
  {The identification rule of $\gerh \cong \ger{u}(1)$
and $\ger{s}\ger{o}(2)$
   we are choosing is
$$
  \frac{i}{2}t ~ \llerarr ~
\left( \ba{rr}
         0 & 1 \\
        -1 & 0  \ea \right). $$},
the partition function \eqn{gwzw model}
can be rewritten as follows;
\be
Z = \int \ca{D}a \ca{D}\bm{g}\, Z_{\msc{gauge}}\lb a,\bm{g}\rb
    \equiv  \int \ca{D}\ca{A} \, Z_{\msc{gauge}} \lb \ca{A} \rb  .
\ee
Notice that,
as we saw in the previous section,
the partition function of the matter $Z_{\msc{gauge}} \lb \ca{A} \rb$
is topologically invariant.
Namely it is
invariant (modulo BRST) under any infintesimal variation of $\ca{A}$;
\bea
- \frac{1}{Z_{\msc{gauge}} \lb \ca{A} \rb}
\frac{\delta Z_{\msc{gauge}} \lb \ca{A} \rb}{\delta \ca{A}} &=&
  - \frac{1}{Z_{\msc{gauge}} \lb \ca{A} \rb}
\left( \frac{\delta Z_{\msc{gauge}} \lb \ca{A} \rb}{\delta a_z}, ~
  \frac{\delta Z_{\msc{gauge}} \lb \ca{A} \rb}{\delta a_{\bar{z}}}, ~
 \frac{\delta Z_{\msc{gauge}} \lb \ca{A} \rb}{\delta e_z^+},~
 \frac{\delta Z_{\msc{gauge}} \lb \ca{A} \rb}{\delta e_{\bar{z}}^-}
\right)   \nonumber \\
& \sim & \left\langle \left\{ Q_{\msc{tot}} + \bar{Q}_{\msc{tot}},~
   (\al_+ \zeta_z,~ \al_+ \bar{\zeta}_{\bar{z}},~ G^-_{G/H}+ G^-_{H^{\bc}}
    , ~ \bar{G}^-_{G/H}+ \bar{G}^-_{H^{\bc}} )  \right\} \right\rangle.
\nonumber \\
{}~& ~& ~
\eea
So we can perform an additional
gauge fixing for the residual degrees of freedom
according to the standard techniques
of  topological gauge theory \cite{TFT} \cite{TFT2}.
The BRST transformations for
this topological symmetry are defined as follows;
\be
\delta_s \ca{A} = \eta,~~~~ \delta_s \la =  \pi, ~~~~
\delta_s \eta = \delta_s \pi =0,
\ee
where we introduce  ghosts
 $\eta = (\eta^0, \eta^+, \eta^-)$,
anti-ghosts
 $\la = (\la_0, \la_+, \la_- )$,
and $B$-fields
 $\pi = (\pi_0, \pi_+ , \pi_-)$.

Then we will determine  the gauge condition
for this topological symmtry.
It will select the $U(1)$-bundle in which the $U(1)$-gauge field $a$
lives. As we will see in the following the most convenient
gauge condition is\footnote
   {In the following,
    we assume that the $U(1)$-bundle we are working on is
    the unitary frame bundle defined by
    $\bm{g}$ (or $(e^+, e^-)$).};
\be
\ba{l}
  a =  \om(\bm{g})~ (\Llerarr ~ D_a e^+ = D_a e^- =0)~~~
   \mbox{(the torsion free condition)}, \\
  F(a) (\equiv da)  = F_0  ~~~
\mbox{(the gauge condition for the Weyl symmetry)},
\ea
\label{gauge cond for top gravity}
\ee
where $\om(\bm{g})$ is the Levi-Civita connection (as was already introduced)
and $F_0 $ is a fixed real 2-form.
It should be remarked that this gauge condition implies the torsion
free condition and the gauge condition for the Weyl symmetry.
With this gauge condition
we obtain the total gauge fixed form of the model;
\bea
Z_{\msc{tot}} &=& \int \ca{D}\ca{A}\, Z_{\msc{gauge}} \lb \ca{A} \rb \,
    \exp \left\lb- \frac{1}{2\pi i} \int_{\Sigma}
       \delta_s \{ \la_0 (da-F_0) + \la_+ D_a e^+ + \la_- D_a e^- \}
         \right \rb   \nonumber \\
&=& \int \ca{D}\ca{A}\, Z_{\msc{gauge}} \lb \ca{A} \rb
           \, \exp \{ - S^{VV} (\ca{A})\} . \label{ztot}
\eea
Here $S^{VV} (\ca{A})$ is no other than the Verlinde-Verlinde's
(1st step) gauge fixed action
of the pure topological gravity \cite{VV};
\be
\ba{lll}
S^{VV} (\ca{A} )& =& \dsp \frac{1}{2\pi i}
          \int_{\Sigma} \{ \pi_0 (da- F_0) + \pi_+ D_a e^+
           + \pi_- D_a e^- \}  \\
      & & \dsp ~~~+ \frac{1}{2\pi i} \int_{\Sigma} (\la_0 \, d \eta^0 +
       \la_+ D_{\ca{A}} \eta^+ + \la_- D_{\ca{A}} \eta- ),
    \label{VV action}  \\
( D_{\ca{A}}\eta^+ & = & d \eta^+ + ia \wedge \eta^+ -i\eta^0 \wedge e^+, \\
D_{\ca{A}}\eta^- & = & d \eta^- - ia \wedge \eta^- + i\eta^0 \wedge e^-. )
\ea
\ee
{}From the look of \eqn{ztot} {\em we can conclude
that our model can be naturally identified with the K.Li's
model of  topological matter
coupled with  topological gravity!! \cite{KLi}}
(Of course, we must further  fix the residual gauge symmetries
in \eqn{ztot}, i.e.
the symmetries of the diffeomorphisms and the local Lorentz, in order to
accomplish the gauge fixing.
(See \cite{VV} for the detail of this procedure.))

One may be afraid that the above gauge condition
\eqn{gauge cond for top gravity}
might affect the topological invariance of the matter part,
because under this gauge condition
the back-ground gauge field $a$ is no longer independent of
the metric.
However, our case is not so,
because  this effect only adds
 BRST-exact terms to the matter part.
In fact, the total EM tensor $T_{\msc{tot}}$ \eqn{ttotal}
is deformed into $T'_{\msc{tot}}$
by this effect;
\be
T'_{\msc{tot}} =  T_{\msc{tot}} +
  \partial_z J^0_{\msc{tot}} ,
\ee
but it is  equal to the original one modulo BRST-exact terms
because of \eqn{jhtot brst};
\be
T'_{\msc{tot}}  = T_{\msc{tot}} + \{ Q_{H^{\bc}},
                    \al_+ \partial_z \zeta_z  \}.
\ee
So this effect will not affect the topological invariance
of the matter system.

The coupling of our  topological matter and  topological gravity is
 realized as the well-known form \cite{KLi}. That is,
the physical observables are nothing but the tensor products
of the chiral primary fields \eqn{physical observables}
(with some suitable shift of the scalar field $\hat{\vphi}$, see \cite{KLi})
and the observables of the pure topological
gravity \eqn{VV action}.

{}~

\subsection{Relations with the KPZ and the DDK Theories}

\indent
{}~~~~We shall try to take   alternative gauge fixing procedures.
Substantially, imposing the gauge condition
\eqn{gauge cond for top gravity}
is equivalent to inserting
the following identity into \eqn{gwzw model}\footnote
    {Here the path-integration \eqn{gwzw model} are performed
       with the domain of the gauge field restricted to the chiral gauge
       orbit of $ \om (\bm{g})$.};
\be
1 = \int  \ca{D}X \ca{D}Y
\delta (A- {}^h \! \om(\bm{g}))
      \De_{\msc{FP}} , ~~~(h = e^{X+ i Y}).
\label{gauge cond for KPZ DDK}
\ee
This leads to
\be
Z = \int \ca{D}\bm{g}\, Z_{\msc{matter}} \lb \bm{g} \rb ,
\ee
where $Z_{\msc{matter}} \lb \bm{g} \rb$ is given by;
\bea
Z_{\msc{matter}} \lb \bm{g} \rb & =  &  \int \ca{D}(g, \chi, \bar{\chi},
    \psi, \bar{\psi} ) \ca{D}X \ca{D}Y \Delta_{\msc{FP}}  \nonumber \\
    & & ~~~\times \exp \left \lb
      -kS_G (g, \, {}^{e^X}\! \om( \bm{g}))
        -S_{\chi \psi}(\chi, \bar{\chi}, \psi, \bar{\psi} ;\,
            {}^{e^X}\! \om(\bm{g}))\right \rb.
\label{alt gauge fixing 1}
\eea
Observing the simple fact;
\be
  {}^{e^X}\! \om(\bm{g}) =
          \om( e^X \bm{g}) ,
\label{weyl axial}
\ee
we find that we may  integrate the mode of the anomaly of the matter
{\em as the Virasoro anomaly (the Weyl or the gravitational anomaly)
instead of the chiral anomaly\/}. Namely  we may perform
the   path-integration of the metric $\bm{g}$ before completing
the path-integration of the gauge field. For this aim,
we must define  $Z_{\msc{matter}} \lb \bm{g} \rb$
as the functional of the metric $\bm{g}$.
It is a non-trivial ploblem of  renormalization,
because now $Z_{\msc{matter}} \lb \bm{g} \rb $
is still a non-topological matter with $c\neq 0$ before completing
the path-integration of the gauge field,  contrary to
the previous gauge choice
in which the matter $Z_{\msc{gauge}} \lb \bm{g}, a \rb$
\eqn{2nd step gauge fixing} was already
topological. There are two natural choices
of the renormalization condition, i.e.
``the KPZ type''
(renormalize preserving the Weyl invariance) \cite{KPZ} and
``the DDK type'' \cite{DDK}
(renormalize preserving the diffeomorphism invariance).

If we choose the renormalization condition of the KPZ type,
we can absorb the $X$-dependence of the action of \eqn{alt gauge fixing 1}
into the integration measure of the metric $\bm{g}$ making use of the relation
\eqn{weyl axial}.
Namely we may transform the integration variable from
$\bm{g}$ to $ e^X  \bm{g}$ because of the Weyl invariance.
Hence we can drop the gauge volume $\dsp \int \ca{D}X \ca{D}Y
 \Delta_{FP}$ off.
It should be remarked the following fact:
 In the previous discussion we dropped
the integral $\dsp \int \ca{D}Y$ independently,
since this measure has no chiral anomaly.
However, in this case
we must drop them  with  this combination preserved, because
 now we would like to integrate the mode of the Virasoro anomaly
 and so we must always  drop  the gauge volume with the combination of
$c=0$ preserved.

{}From the above discussions
we obtain the following gauge fixed form of the model;
\be
Z = \int \ca{D}\bm{g} \, Z_{\msc{matter}} \lb \bm{g} \rb^{\msc{W}} ,
\label{alt gauge fixing0}
\ee
where
\bea
Z_{\msc{matter}} \lb \bm{g} \rb^{\msc{W}}
& =  &  \int \ca{D}(g, \chi, \bar{\chi},
    \psi, \bar{\psi} )^{\msc{W}} \nonumber \\
& & ~~~~~~~\times  \exp \left \lb -kS_G (g, \om(\bm{g}))
        -S_{\chi \psi}
     (\chi, \bar{\chi}, \psi, \bar{\psi} ;\, \om(\bm{g})) \right \rb.
\label{alt gauge fixing}
\eea
The superscript ``W'' indicates the Weyl invariant definition.
Remark that, in this situation, the gauge fixed action
\eqn{alt gauge fixing} has an  extra dependence   on the metric $\bm{g}$
through the Levi-Civita connection $ \om(\bm{g})$.
Taking account of this fact,
the EM tensor for the matter sector should  get
the following forms;
\bea
T^{\msc{matter}} &=& T_g + \partial_z J_g^0
  + T_{\chi \psi}+ \partial_z J_{\chi \psi}^0
           \label{tmatter}   \\
  & = & T_{(1,p)}^{\msc{FF}} + T_{G/H} ,
\eea
where we set
\bea
T_{(1,p)}^{\msc{FF}} & = & - :(\partial_z \vphi )^2 : +
                          i \al_0 \partial_z ^2 \vphi ,\\
& & ~~~(\al_0 = \al_+ + \al_- \equiv \frac{p-1}{\sqrt{p}},
{}~ p = k+2). \nonumber
\eea
$T_{(1,p)}^{\msc{FF}}$ is no other than the well-known Feigin-Fuchs
representation of the $(1,p)$-conformal matter with  central charge
$\dsp c_{1,p} = 1 - \frac{6(p-1)^2}{p}\, $ \cite{DF}!!
In fact the ``twist'' in the expression
of the RHS of \eqn{tmatter} completely
coincides with that of
the quantum Hamiltonian reduction {\em \'{a}la\/}
Drinfeld-Sokolov \cite{BO}.
Remembering the BRST-exactness of $T_{G/H}$,
which is indeed the EM tensor of the KS sector
we are  intending to kill out, we can conclude
that {\em our model is essentially equivalent to
the $(1,p)$-conformal matter coupled to gravity, with the correspondence;
$p= k+2$.}

To complete the quantization we must perform
the residual integral, i.e.
of the metric $\bm{g}$.
Since we have renormalized the matter part
with the Weyl invariance preserved,
we necessarily suffer the gravitational anomaly;
\be
  Z_{\msc{matter}} \lb {}^f \! \bm{g} \rb^{\msc{W}}
   = Z_{\msc{matter}} \lb \bm{g} \rb^{\msc{W}} \, \exp
        \left \{ \frac{c_{1,p}}{24}
            S_{\msc{KPZ}} ( f ;\, \bm{g} )  \right\} ,
\label{gravitational anomaly}
\ee
where $f$  is an arbitrary diffeomorphism on $\Sigma $ and
${}^f \! \bm{g} \equiv f^{-1*} \bm{g}$.
The ``KPZ action'' \cite{KPZ}
in \eqn{gravitational anomaly}
 is defined as follows\footnote{This definition is  the same
    as the well-known one.
    If we take the ``light-cone gauge''
    (on the 2D space-time with the metric of the Lorentzian signature),
    our definition \eqn{KPZ action} reduces to the form given in \cite{KPZ}.};
\be
S_{\msc{KPZ}} ( f ;\, \bm{g} )  =  \frac{1}{2\pi}
\int_{\Sigma} dv({}^f\! \bm{g})\, \langle ({}^f\! \bm{g})^* ,
       \ca{S} (f^{-1}) \rangle \label{KPZ action}
\ee
where
\bea
\ca{S} (F) & =& \ca{S}_{zz}(F)  dz \!\otimes \! dz
  + \ca{S}_{\bar{z}\bar{z}}(F) d\bar{z} \! \otimes \! d\bar{z}, \nonumber \\
\ca{S}_{zz}(F) & \equiv & \frac{F_{zzz}}{F_z} - \frac{3}{2}
   \left( \frac{F_{zz}}{F_z}  \right)^2  ,
  ~~~ \ca{S}_{\bar{z}\bar{z}}(F) \equiv \overline{\ca{S}_{zz}(F)} .
\eea
In the above definition ``$dv(\bm{g})$'' denotes the
canonical volume element defined by the metric $\bm{g}$, ``$\bm{g}^*$''
denotes the dual metric of $\bm{g}$, and ``$\langle ~,~ \rangle$''
means the usual contraction of tensor fields.

Needless to say, this KPZ action is related with  the gravitational anomaly
as the gauged WZW action \eqn{gwzw action}
and the Liouville action are related with
the chiral and the Weyl anomalies respectively.
Here the Liouville action is defined as \cite{DDK}, \cite{Gervais}
(with the cosmological term omitted);
\be
S_{\msc{L}} ( \sigma ; \bm{g} ) =
\frac{1}{2\pi i} \int_{\Sigma} \{ \deebar \sigma \partial \sigma +
   2 i R(\bm{g}) \sigma \}.
\label{Liouville action}
\ee
The RHS \eqn{Liouville action}
is defined with respect to  the complex structure
so that the metric $\bm{g}$ becomes K\"{a}hler.
The KPZ action is manifestly Weyl invariant, but behaves as the
1-cocycle with respect to diffeomorphisms.
On the contrary, the Liouville action
is clearly invariant under any diffeomorphism, but is the 1-cocycle
with respect to the Weyl rescaling.

In order to integrate the metric $\bm{g}$, we need to clarify
the definition of the measure $\ca{D} \bm{g}$
(the measure of the ``dynamical metric'').
It should be remarked that {\em the measure $\ca{D}\bm{g}$ should
possess no Virasoro anomaly, because, if this is not the case,
it is contrary to the existence of topological gravity.}
Therefore,
if we parametrize  the dynamical metric $\bm{g}$ as
$\bm{g} = e^{\sigma} ({}^f \! \hat{\bm{g}}) $,
where $e^{\sigma}$, $f$ of course mean
respectively the ``Liouville mode'', the mode of the diffeomorphisms,
and $\hat{\bm{g}}$
is the ``back-ground metric'' (the modulus of $\bm{g}$),
we must {\em define\/} this measure as follows;
\bea
\ca{D} \bm{g} & = &
\ca{D}\hat{\bm{g}}  \, \ca{D}(\sigma ;\, \hat{\bm{g}}) \,
\exp \left\{\frac{1}{24}S_{\msc{L}} ( \sigma ; \, \hat{\bm{g}}) \right \} \,
\ca{D}(f ;\, \hat{\bm{g}}) \nonumber \\
  & & ~~~~~\times \exp \left \{
   -\frac{26}{24}S_{\msc{KPZ}}( f; \hat{\bm{g}} )\right\}
       \Delta_{\msc{FP}}(\hat{\bm{g}}) ^{\msc{W}}.
\eea
Here $\ca{D} \hat{\bm{g}}$, $\ca{D}(\sigma ;\, \hat{\bm{g}})$,
$\ca{D}(f ;\, \hat{\bm{g}})$
denote respectively the measure of the modulus,
that of the Liouville mode, and that of the mode of diffeomorphisms,
which are defined associated with the back-ground metric $\hat{\bm{g}}$.
$  \Delta_{\msc{FP}} (\hat{\bm{g}}) ^{\msc{W}} $ is
the usual FP determinant (i.e. the Jacobian between the functional
measure $\ca{D}(f ;\, \hat{\bm{g}})$ and $\ca{D}{}^f\! \hat{\bm{g}}$)
renormalized to be Weyl invariant.
In fact this definition  has no anomaly:
The measure of the modulus $\ca{D} \hat{\bm{g}}$, which is nothing but
a finite dimensional measure, does not contribute, and
$\ca{D}(f ;\, \hat{\bm{g}})$  has clearly no anomaly.
Moreover $\ca{D}(\sigma ;\, \hat{\bm{g}})$ and
$\Delta_{\msc{FP}}(\hat{\bm{g}})^{\msc{W}}$
possess respectively $c=1$, $c=-26$, which are precisely
canceled with the (classical) anomalies in the functionals
$\dsp \exp \left\{ \frac{1}{24}S_{\msc{L}}
( \sigma ; \, \hat{\bm{g}}) \right \}$ ($c=-1$)
and $\dsp \exp \left \{
   -\frac{26}{24}S_{\msc{KPZ}}( f; \hat{\bm{g}} )\right\}$ ($c=26$).
(c.f. \cite{DDK}, \cite{KPZ}, \cite{BO})
Alternatively we may say that the following identity should hold;
\bea
1 &=& \int \ca{D}\hat{\bm{g}}  \, \ca{D}(\sigma ;\, \hat{\bm{g}}) \,
\exp \left\{\frac{1}{24}S_{\msc{L}} ( \sigma ; \, \hat{\bm{g}}) \right \} \,
\ca{D}(f ;\, \hat{\bm{g}}) \nonumber \\
  & & ~~~~~\times \exp \left \{
   -\frac{26}{24}S_{\msc{KPZ}}( f; \hat{\bm{g}} )\right\}
      \Delta_{\msc{FP}}(\hat{\bm{g}})^{\msc{W}} \,
   \delta(\bm{g} - e^{\sigma}({}^f\! \hat{\bm{g}})) .
\eea

Inserting this identity into \eqn{alt gauge fixing0},
we obtain the final gauge fixed form;
\be
Z = \int \ca{D}\hat{\bm{g}} \,
  \ca{D}(f,b,\bar{b},c,\bar{c};\, \hat{\bm{g}})^{\msc{W}}\,
   Z_{\msc{matter}} \lb \hat{\bm{g}} \rb^{\msc{W}}\,
\exp \left\{ -\frac{26-c_{1,p}}{24} S_{\msc{KPZ}} ( f ; \, \hat{\bm{g}} )
-S_{bc}(b ,\bar{b},c,\bar{c} ; \, \hat{\bm{g}}) \right\} .
\ee
Here we have introduced the usual $b,c$-ghosts with $c=-26$
to rewrite the functional determinant
$ \Delta_{\msc{FP}} (\hat{\bm{g}})^{\msc{W}}$.
This implies that by choosing the renormalization condition
of the ``KPZ type'', our model reproduces the famous formulation
of the 2D induced gravity by Knizhnik-Polyakov-Zamolodchikov \cite{KPZ}.

The total central charge is of course equal to zero \cite{KPZ}.
\be
c^{\msc{tot}}=c_{1,p} + c_f + c_{bc}
= c_{1,p} + (26 -c_{1,p}) +(-26) =0.
\ee

{}~

Next let us  turn to the case  that
the  renormalization condition of the DDK type,
i.e. renormalizing without the gravitational anomaly, is chosen.
We shall express the partition function of the matter part under this
renormalization condition by ``$Z_{\msc{matter}} \lb \bm{g} \rb^{\msc{D}}$''.
(The superscript ``D'' indicates the diffeomorphism invariance
corresponding to the superscript ``W'' indicating the Weyl invariance.)
The transformation from $Z_{\msc{matter}} \lb \bm{g} \rb^{\msc{W}}$
to $Z_{\msc{matter}} \lb \bm{g} \rb^{\msc{D}}$
is essentially the well-known procedure of the ``finite renormalization''
with the bare Lagrangian fixed, and so we can likewise regard the matter
$Z_{\msc{matter}} \lb \bm{g} \rb^{\msc{D}}$ as the $(1,p)$-conformal matter.
However, in this case the effect of
non-zero central charge of the matter part
appears as the Weyl anomaly in place  of the gravitational anomaly;
\be
  Z_{\msc{matter}} \lb e^{\sigma}  \bm{g} \rb^{\msc{D}}
   = Z_{\msc{matter}} \lb \bm{g} \rb^{\msc{D}} \,
         \exp \left \{ \frac{c_{1,p}}{24}
           S_{\msc{L}} ( \sigma ;\, \bm{g} )  \right \}.
\ee

In the similar way as the KPZ type,
inserting into \eqn{alt gauge fixing} the identity;
\bea
1 &=& \int \ca{D}\hat{\bm{g}} \, \ca{D}(\sigma ;\, \hat{\bm{g}}) \,
\exp \left \{ \frac{1}{24}S_{\msc{L}} ( \sigma ;\,  \hat{\bm{g}}) \right \} \,
\ca{D}(f ; \hat{\bm{g}})  \nonumber \\
    &  & ~~~~~~ \times \exp \left\{
    -\frac{26}{24}S_{\msc{L}}( \sigma ;\, \hat{\bm{g}} )  \right \}
        \Delta_{FP} (\hat{\bm{g}}) ^{\msc{D}} \,
   \delta(\bm{g} - e^{\sigma}({}^f\! \hat{\bm{g}})) ,
\eea
we arrive at the final gauge fixed form;
\be
Z = \int \ca{D}\hat{\bm{g}} \,
  \ca{D}(\sigma ,b,\bar{b},c,\bar{c};\, \hat{\bm{g}})^{\msc{D}}\,
   Z_{\msc{matter}} \lb \hat{\bm{g}} \rb^{\msc{D}}\,
\exp \left\{ -\frac{25-c_{1,p}}{24} S_{\msc{L}}(\sigma ;\, \hat{\bm{g}} )
-S_{bc}(b ,\bar{b},c,\bar{c} ;\, \hat{\bm{g}}) \right \}.
\ee
This means that under this ``DDK type'' renormalization condition,
our model reduces to the famous formulation
by David-Distler-Kawai \cite{DDK}.

The total central charge also vanishes \cite{DDK};
\be
c^{\msc{tot}} = c_{1,p} + c_{\sigma} + c_{bc}
  = c_{1,p} +(1 + 24\times \frac{25-c_{1,p}}{24}) + (-26) =0 .
\ee

{}~

\section{Conclusions and Discussions}

\cleqn
\indent
We have quantized the topological gauged WZW model
associated with $SU(2)/U(1)$ by  path-integration.
Our quantization scheme can be immediately generalized to the higher rank
cases, i.e. to the cases for general compact K\"{a}hler homogeneous
spaces $G/H$ (not necessarily hermitian symmetric)
in the almost similar manner.
This will be one of the main subject of the subsequent paper \cite{ns 1}.

We have also shown that the topological $SU(2)/U(1)$-model naturally reduces
to the known theory of 2D gravity with matter. There were three types
of the gauge fixing procedures
which include the designation of the renormalization
condition; the K.Li type, the KPZ type, and the DDK type.
The crucial difference among these gauge conditions consists in
the  choice of the methods to cancel out  the anomaly of the matter
originated in the gauged WZW action, that is,
the choice of whether one integrates
the mode of the chiral anomaly, the mode of the gravitational anomaly,
or the mode of the Weyl anomaly.

In the K.Li gauge we can get the topological matter
before performing the path-integration of the metric,
and hence  automatically make   topological gravity couple with it.
On the contrary,
in the KPZ or the DDK gauge we must integrate the metric in order to
make the central charge of the system  vanish.
It has not yet been known how we should perform the integration
on the moduli space in these gauges.

Let us recall  that in the the K.Li gauge
the physical operators (of the matter part)
are given as the  chiral ring generated by
the elements having ``vertex operator'' forms;
$e^{i  \beta \hat{\vphi}(z)}$,
where $\beta $ takes a value in
$\{ ~(1-s)\al_- ~;~s =1,\cdots ,k+1 \equiv p-1~\}$.
This can be rewritten in the form
$e^{ i \beta \vphi(z)}\, e^{\beta X(z)}$, and the ``$\vphi$-part''
$e^{i  \beta \vphi(z)}$ is no other than one of the  primary fields
of the $(1,p)$-conformal matter. So we can naturally consider the chiral
primary fields in the K.Li gauge  to be
the ``primary fields dressed with  gauge field
(or gravity)'' of the conformal matter.

There are similar situations in the KPZ or the DDK gauge, too.
In fact, it is familiar that
in the DDK theory any  primary field is  dressed with
the Liouville field $\sigma$ so that the total conformal weight
is equal to $(1,1)$ in the influence of gravity \cite{DDK}.
In the KPZ theory the appearance of the gravity dressing is not so manifest,
since this effect  is described
by the ``hidden $SL(2, \br)$-current algebra'', which can be
understood by making related to the  coadjoint orbits of the Virasoro
group (the group of  diffeomorphisms on $S^1$ with  central extension),
or from the viewpoint of the Hamiltonian reduction
of the $SL(2,\br)$-WZW model. \cite{AS} \cite{BO}
But this is thought  to  have intrinsically
 the same origin as above.

It is well-known that
 topological gravity is  equivalent to
the intersection theory on the moduli space of $\Sigma$ \cite{witteng}.
The physical observables (usually denoted by ``$\sigma_n$'')
are identified with the de Rham cohomology classes on the moduli space.
The degrees of freedom for  infinitesimal deformations
of the modulus are carried by  zero-modes of the $\eta$-ghost
(it is in our notation, of course it is written
as ``$\psi$'' in \cite{VV}), which are identified
with 1-forms on the moduli space.

On the other hand, in the KPZ or the DDK gauge
the degrees of freedom for the deformations of the modulus
are carried by the zero-modes of the $b$-(anti-)ghost,
which  correspond to  vector fields on the moduli space.
The correspondence of the physical observables between
the K.Li gauge and the KPZ or the DDK gauge is only
a partially solved problem even now \cite{LZ}.

It is also  intereting to search  the correspondence between
these theories for the general $(q,p)$-conformal matter.
It might be  suggestive that in our formalism
the $(1,p)$-conformal matter comes into sight
by suffering the same twist which occurs in the quantum
Hamiltonian reduction {\em \'{a}la\/} Drinfeld-Sokolov \cite{BO}.
So naively it might be plausible that the correspondence can be extended
to the case of  the general $(q,p)$-matter such as $\dsp k+2 = \frac{p}{q}$,
since this correspondence is
the well-known one in the theory of the quantum Hamiltonian reduction.
In topological gravity
 we must suitably place
some physical operators as the back-ground sources
in order to treat the general $(q,p)$-matter \cite{Distler},
while,
in the free field realization of the WZW model \cite{Wakimoto},
the parameter $k$ appears as the back-ground charge of the model.
Hence there might exist some kind of
connection between changing the parameter $q$
and making the level $k$ fractional.

{}~

The extensions of our investigation on
 2D gravity to  the cases of  general flag manifolds $G/H$
are also  interesting subjects.
The quantum theory of these cases will include
``topological Yang-Mills'' besides  topological gravity.
In other words  we may
say that in these systems
gravity and  gauge fields coexist and
both couple  with some conformal matters
(a continuum limit of the ``lattice gauge  theory on the random lattice''!).
Especially we can make the  following observation in the topological
$G/T$-model ($T$ is the maximal torus of $G$). \cite{ns 1}:
If we quantize this model in the same way as the $SU(2)/U(1)$-case
with the  KPZ or the DDK like gauge,
we find out
that the model suffers the same twist as the Hamiltonian reduction
\cite{toda} as above.
{}From this observation it is plausible that we can get
the $\mbox{W}G$-conformal matter from the topological $G/T$-model.
Therefore it will be meaningful to study
the ``$\mbox{W}G$-gravity'' \cite{W gravity1}, \cite{W gravity2}
based on the topological $G/T$-model.
The detail will be  given in the subsequent paper \cite{ns 1}.

{}~

As the last remark of this paper,
we comment on the phenomenon that happens when
we set the back-ground gauge field $a$ as $a = \om_{\msc{spin}}(\bm{g})$
(``the spin connection'', symbolically we can write it as
``$\dsp \frac{1}{2} \om (\bm{g})$''),
instead of fixing as  $a = \om(\bm{g})$.
In this situation we find that the KS sector of the model
suffer the ``inverse twist'',
and we can regain the (untwistd) $N=2$ KS model
for $\bc P^1$,
which is a theory of $c \neq 0$.
But the CG sector is not affected, since the back-ground
gauge field $a$ is decoupled from the ghosts of the CG sector
$\xi$, $\zeta$.

The similar  phenomenon exists in the  higher rank cases, too.
In the case $G/H$ is hermitian symmetric we can get the corresponding
KS model. But in a generic case,
although we can easily show that the KS sector becomes
some non-topological CFT as well,
we cannot reproduce the original supersymmetric gauged WZW model,
especially
it is not likely that there  exists  a $N=2$ SUSY.

These  studies will be given in the paper \cite{ns 2}.

\newpage

\appendix
{\Large \bf Appendix}
\section{Another Derivation of the Topological Gauged WZW Model}
\cleqn

In this appendix we intend to derive the topological $G/H$-model,
which was defined as the twisted version of the supersymmetric
gauged WZW model \eqn{susy gwzw},
from the $G/G$-gauged WZW model.
The $G/G$-gauged WZW model is nothing but the one regarding the gauge
field $A$ as $\gerg$-valued in \eqn{gwzw action};
\be
Z_{G/G} = \int \ca{D}\bm{g} \ca{D}g \ca{D}A  \,
\exp \{ -kS_G(g,A) \}
\equiv \int \ca{D}\bm{g} \, Z_{G/G} \lb \bm{g} \rb .
\label{gwzw model app}
\ee
Of course this model is equivalent to the $G/G$-coset CFT,
which is obviously a theory of $c=0$, so naively
a trivial theory with no physical degrees of freedom.\footnote
    {More strictly, we can find that this model becomes  equivalent
       to  topological gravity plus topological Yang-Mills
      $\dsp Z = \int \ca{D}\bm{g} \ca{D}A$ \cite{G/G}.}
As was already commented, the mechanism of this type cancellation
of central charge can be understood by means of the similar logic given in
\cite{DDK}. The path-integrations of ``geometrical fields'',
such as metric (or complex structure), gauge fields, are capable to
cancel the anomalies of the matter coupled with them.
Particularly, in the $G/H$-gauged WZW model such cancellation of the chiral
anomaly is ``partial'', while, in the $G/G$-case the cancellation occurs
completely.
Since the  model we want is topologically invariant but has
non-trivial finite physical degrees of freedom, we must try to avoid
this triviality.

Now we shall give the following simple observation;
Let $H$ be a closed subgroup of $G$
such that the coset space $G/H$ becomes K\"{a}hler, which implies
in particular that $H$ includes the maximal torus $T$ of $G$.
 We denote its Lie subalgebra by $\gerh$.
In this situation any element $g$ of $G$ can be expressed as
$ g = {}^f\! h$, $h \in H$, $f \in G$, since  $H$ includes the
maximal torus. Making use of this fact pointwisely,
we can connect  necessarily
the chiral field $g$ to some $H$-valued field along
a vectorial (so, non-anomalous) gauge orbit.
Hence we may suppose that
the chiral anomaly of the gauged WZW action in effect
exists only along the (axial) $H^{\bc}$-direction.
Taking this observation into account, it is plausible that
{\em the $G/G$-gauged WZW model leaves the topological invariance
(i.e. the property $c=0$), even if we restrict the domain of the
path-integration of  $A$ to the subspace composed of the elements
expressible in the form $A =  {}^{\Om}\! A'$, where $\Om$ is
a $G$-valued gauge transformation and $A'$ has only $\gerh$-components.}
In fact, from the above observation
 the integral of the gauge field $A$ so restricted is sufficient
to cancel out the chiral anomaly.

In the following arguments we shall impose this restriction
on the functional space of $A$ integrated,
that is, define the path-integration space of
$A$ as follows;
\be
\ba{l}
\gerA_{\msc{rest}} = \{ ~ A~:~ A= {}^{\Om}\!({}^h \! a),~
\forall \Om \in \ca{G}_{G},~ \forall h \in \ca{G}_H^{\bc}, \\
{}~~~~~~~~~~~ a~:~
\mbox{the modulus of the gauge field} \},
\ea
\label{restricted space}
\ee
where $\ca{G}_G$ denotes the space of
$G$-valued (vectorial) gauge transformations
and $\ca{G}_H^{\bc}$ denotes the space of $H^{\bc}$-valued chiral
gauge transformations.

Let us try to perform the gauge fixing of the model
\eqn{gwzw model app} for the $G/H$-part,
keeping the above assumption in mind.
We shall follow the standard BRST prescription, with the gauge condition;
\be
 ( \germ_+, A^{10}) =0,~~~(\germ_-, A^{01})=0.
\label{gauge cond app}
\ee
($\germ_{\pm}$ was introduced in section 2, see \eqn{parabolic}.)
Actually it is sufficient to impose only the 1st condition,
since our gauge field $A$ has the property
$A^{10\dag} = -A^{01}$.
We introduce the following BRST ghost system and the corresponding
BRST transformations;
\be
\ba{l}
\mbox{ghosts}~~~~
\psi~:~ \mbox{$\germ_+$-valued (0,0)-form} ,~~~
\bar{\psi} \equiv \psi^{\dag} ~:~ \mbox{$ \germ_-$-valued (0,0)-form}, \\
\mbox{anti-ghosts}~~~~
\chi~:~ \mbox{$\germ_-$-valued (1,0)-form},~~~
\bar{\chi} \equiv \chi^{\dag}~:~ \mbox{$\germ_+$-valued (0,1)-form}, \\
\mbox{$B$-fields}~~~~
B ~:~ \mbox{$\germ_-$-valued (1,0)-form},~~~
\bar{B} \equiv B^{\dag} ~:~\mbox{$\germ_+$-valued (0,1)-form}, \\
\ba{ll}
\mbox{BRST transformations}
& \delta_{G/H} A =  D_A \psi \equiv d \psi + \lb A, \psi \rb ,\\
&\bar{\delta}_{G/H} A =  D_A \bar{\psi} \equiv d \bar{\psi}
      + \lb A, \bar{\psi} \rb ,\\
&\delta_{G/H} g = \lb \psi , g \rb, ~~~
\bar{\delta}_{G/H} g = \lb \bar{\psi} , g \rb, \\
&\delta_{G/H} \chi = B,~~~
\bar{\delta}_{G/H} \bar{\chi} = \bar{B},\\
&\dsp \delta_{G/H} \psi = \frac{1}{2} \lb \psi, \psi \rb ,~~~
\dsp \bar{\delta}_{G/H} \bar{\psi} =
     \frac{1}{2} \lb \bar{\psi}, \bar{\psi} \rb ,\\
&\dsp \delta_{G/H} \bar{\psi} = \bar{\delta}_{G/H} \psi
   = \frac{1}{2} \lb \psi, \bar{\psi} \rb ,~~~ \\
&\mbox{(other combinations are defined to vanish)}.
\ea
\ea
\label{g/h brst}
\ee
We obtain the partition function with
the  gauge fixing as follows;
\bea
Z_{G/G,~ \msc{gauge}} \lb \bm{g} \rb
&=& \int \ca{D} (g, A , B,\bar{B}, \chi, \bar{\chi}, \psi, \bar{\psi})
 \nonumber \\
& & ~~~~~~~~\times  \exp \left\lb -kS_G(g,A) - \frac{1}{2\pi i}
\int (\delta_{G/H}+ \bar{\delta}_{G/H}) (\chi - \bar{\chi},A ) \right\rb \\
  & = & \int \ca{D} (g,A', \chi, \bar{\chi}, \psi, \bar{\psi} ) \nonumber\\
 & & ~~~~~~~ \times  \exp \left \lb -kS_G(g,A') - \frac{1}{2 \pi i} \int
  \{ (\deebar_{A'}\psi, \chi )-(\bar{\chi}, \partial_{A'}\bar{\psi})\}
\right\rb,
\label{gauge fixing app}
\eea
where  $A'$ has only $\gerh$-components.
In the 2nd line of the above equation
we have integrated the $B$-fields out,
which impose on the gauge field $A$ the constraints
\eqn{gauge cond app}.
Solving this constraints on the restricted space
 \eqn{restricted space}, we obtain the gauge field
$A'$ possessing only $\gerh$-components.
In fact, since we have already used the gauge degrees of freedom
of the $G/H$-part, $A'$ cannot have the $G/H$-components by the
assumption; $A \in \gerA_{\msc{rest}}$.
However, strictly speaking, this consideration is somewhat too naive.
Actually
there still exist the residual gauge degrees of freedom of the $G/H$-part
which correspond to the zero-modes of the ghosts $\psi$, $\bar{\psi}$.
But to make things easy, we shall neglect these zero-modes
for the time being.

{}From the look on \eqn{gauge fixing app},
we find that it can be identified with  the topological gauged WZW model
associated with $G/H$ defined by twisting
the corresponding supersymmetric model in section 2!!
We can further show that
the BRST symmetry \eqn{susy brst}
  introduced in section 2,
which was originated in the N=2 SUSY of the untwisted model,
is thought to be  the on-shell counterparts of the ``off-shell''
BRST symmetry for the $G/H$-gauge
transformations \eqn{g/h brst}.
In this respect we should notice the following fact:
In the standard local operator formalism of the WZW model,
the Hilbert space is factorized
into the holomorphic and the anti-holomorphic sectors;
$\ca{H} = \ca{H}^+ \otimes \ca{H}^-$, and
the right (left) action of
the loop group on $\ca{H}^+$ (resp. $\ca{H}^-$)
is defined to be trivial.

In the $SU(2)/U(1)$-case,
as was already shown in section 2, after performing
the residual path-integration of the model
(i.e. $H^{\bc}$-part or the CG part),
we indeed get a topological conformal model with non-trivial
(finite) physical degrees of freedom, that is, the chiral primary ring.
In general cases we can get the similar results, too.
This fact will be shown in the paper \cite{ns 1}.

The above derivation of the topological gauged WZW model may be
somewhat technical. But it is expected
to give a transparent interpretation
as regards the geometrical back-grounds of the model, especially
of the KS sector.
Now it is clear that the moduli space of the KS sector
is indeed the K\"{a}hler space $G/H$, which corresponds to
the space of constant modes of gauge transformations.
At this stage,
 we cannot neglect the zero-modes
of the ghost fields $\psi$, $\bar{\psi}$.
The bosonic partners of these zero-modes
must exist in both the space of the gauge field $A$
and the space of the chiral field $g$.
For $A$, this degrees of freedom of modulus is no other than
the ``residual components of $G/H$-part''
omitted in the above discussion.
While, the modulus  of $g$ is considered as
generated by the action of  zero-modes
of the $\germ_{\pm}$-components of the chiral current of $g$.
Of course these degrees of freedom are at most finite dimensional,
so  do not affect in essence the above arguments
of the derivation.
They  may correspond to the holomorphic and anti-holomorphic coordinates
on the moduli space, i.e. on
the K\"{a}hler manifold $G/H$.
Correspondingly the zero-modes of ghosts $\psi$, $\bar{\psi}$ are identified
with the (1,0)- and (0,1)-forms on this space,
which was already mentioned in section 2, and will
generate the Dolbeault cohomology algebra on $G/H$ (c.f. \cite{LVW}).

{}~

{}~

\section*{Acknowledgements}

We would like to thank
Prof.T.Eguchi for several useful discussions and continuous
encouragement during which this work was brought to fruition.
We also thank Dr.S.Hosono and Dr.H.Kunitomo for the helpful discussions on
topological gravity.
T.N would further like to thank Prof.Y.Kazama
for the useful comments on this work.

\newpage


\end{document}